\documentclass[aps,twocolumn,groupedaddress,amsmath,amssymb,nofootinbib]{revtex4}
\usepackage{graphicx}
\usepackage{dcolumn}
\usepackage{bm}
\usepackage{amssymb}
\usepackage{amsmath}
\usepackage{epsfig}    
\usepackage{color}
\usepackage{slashed}
\usepackage{hhline}

\def\be{\begin{equation}}
\def\ee{\end{equation}}
\newcommand{\bea}{\begin{eqnarray}}
\newcommand{\eea}{\end{eqnarray}}
\newcommand{\nn}{\nonumber}

\begin{document}


\title{ 
 Explaining $B\to K^{(*)}\ell^+ \ell^-$ anomaly \\ by radiatively induced coupling 
in $U(1)_{\mu-\tau}$ gauge symmetry}

\author{ P. Ko}
\email{pko@kias.re.kr}
\affiliation{School of Physics, KIAS, Seoul 02455, Korea}
\affiliation{Quantum Universe Center, KIAS, Seoul 02455, Korea}

\author{Takaaki Nomura}
\email{nomura@kias.re.kr}
\affiliation{School of Physics, KIAS, Seoul 02455, Korea}

\author{Hiroshi Okada}
\email{macokada3hiroshi@cts.nthu.edu.tw}
\affiliation{Physics Division, National Center for Theoretical Sciences, Hsinchu, Taiwan 300}

\date{\today}

\begin{abstract}
 We propose a scenario to generate flavor violating $Z'$ interactions at one loop level,  
by  introducing
$U(1)_{\mu-\tau}$ gauge symmetry, extra vectorlike quark doublets 
$Q'_a$ and singlet scalar $\chi$.  Both  $Q'_a$ and $\chi$ are charged under  $U(1)_{\mu-\tau}$ and carry 
{odd} dark $Z_2$ parity. 
Assuming that $\chi$ is the dark matter (DM) of the universe and imposing various constraints from 
dark matter search, flavor physics and collider search for $Q'_a$, one can show that radiative corrections to 
$b\rightarrow s Z^{'*} \rightarrow s l^+ l^- $ involving $Q'_a$ and $\chi$ can induce $\Delta C_9 \sim -1$ 
which can resolve the  LHCb anomalies related with $B\to K^{(*)} \ell^+ \ell^-$.  
Therefore both DM and $B$ physics anomalies could be accommodated in the model.
\end{abstract}
\maketitle

\section{Introduction}
Flavor violating interactions via new gauge boson $Z'$ is one of the interesting possible  
physics scenarios beyond the standard model (BSM).
 For the last few years there have been some indication of such interactions in $B$ physics;
the angular observable $P'_5$ in decay of $B$ meson, 
$B\to K^* \mu^+ \mu^-$~\cite{DescotesGenon:2012zf}, where $3.4\sigma$ deviations are measured 
from the integrated luminosity of 3.0 fb$^{-1}$ at the LHCb~\cite{Aaij:2015oid}, confirming an earlier 
result with $3.7\sigma$ deviations~\cite{Aaij:2013qta}. Moreover, $2.1\sigma$ deviations were reported  
in the same observable by Belle~\cite{Abdesselam:2016llu, Wehle:2016yoi}. In addition, an anomaly in the measurement of the ratio $R_K = BR(B^+ \to K^+ \mu^+\mu^-)/BR(B^+ \to K^+  e^+e^-)$~\cite{Hiller:2003js, Bobeth:2007dw} at the LHCb indicates a $2.6\sigma$ deviations from the lepton universality predicted in the SM~\cite{Aaij:2014ora}.  
{ Moreover the LHCb collaboration also presented the ratio $R_{K^*} = BR(B \to K^* \mu^+\mu^-)/BR(B \to K^*  e^+e^-)$ which is deviated from the SM prediction by $\sim 2.4 \sigma$ as $R_{K^*} = 0.660^{+0.110}_{-0.070} \pm 0.024 (0.685^{+0.113}_{-0.069} \pm 0.047)$ for $(2 m_\mu^2) < q^2 < 1.1$ GeV$^2$ (1.1 GeV$^2 < q^2 < 6$ GeV$^2$)~\cite{Aaij:2017vbb}. }
One of the explanations for these anomalies in the $B$ decay could come from $Z'$ 
which has flavor dependent interactions in  the quark sector~\cite{Crivellin:2015lwa, Ko:2017quv, 
Altmannshofer:2015mqa,Boucenna:2016wpr,Boucenna:2016qad,GarciaGarcia:2016nvr,Altmannshofer:2014cfa, Altmannshofer:2016jzy} and 
can induce shift of the Wilson coefficient $C_9$ where  the shift $\Delta C_9 \sim -1$ is indicated to resolve 
the anomalies~\cite{Descotes-Genon:2013wba, Descotes-Genon:2015uva, Hurth:2016fbr,Altmannshofer:2014rta}.
In previous attempts, flavor violating $Z'$ interactions in the quark sector were obtained at tree 
level, assuming non-trivial charge assignments of extra U(1) gauge symmetry or nonzero mixings between 
quarks and new vector-like quarks charged under extra U(1). 
On the other hand,  flavor dependent  couplings can also arise  at loop levels if we add few exotic fermions and/or 
scalar fields.  Furthermore, if these extra particles  have $Z_2$ odd dark parity, 
motivated by dark matter of the universe,  such a scenario provides interesting connection between 
$B$ physics anomaly and DM physics.

In this letter, we propose a new resolution of these $B$ physics anomalies by introducing exotic vector-like 
quarks ($Q'$) and an inert singlet boson ($\chi$) which are charged under the gauged $U(1)_{\mu-\tau}$ 
symmetry and have $Z_2$ odd parity {which guarantees dark matter stability}
~\footnote{In this type of symmetry, some specific textures can be obtained. 
Therefore one can obtain some predictions in the neutrino sector, although we will not discuss here. 
See ref.~\cite{Baek:2015mna} for instance.}.   These two new fields play an crucial role in connecting 
leptons and quarks at one-loop level. Furthermore, $\chi$ is {assumed to be the lightest 
$Z_2$-odd particle, making} the DM candidate within our model. 
We then explore explanation of $B \to K^{(*)} \ell^+ \ell^-$ anomaly and relic density of DM, simultaneously 
taking into account various constraints from the $B_s-\bar B_s$ meson mixing, $b\to s\gamma$, 
and direct detection of DM via $Z'$ portal at one-loop level originating from these new fields.

This letter is organized as follows.
In Sec. II, {we present our model and study $B$ physics and DM phenomenology:  
the Wilson coefficients for $B\to K^{(*)} \ell^+ \ell^-$ anomaly and  $B_s-\bar B_s$ meson mixing, 
the branching ratio of $b\to s\gamma$,  thermal relic density of DM, and the spin independent DM-nucleon 
scattering cross section via $Z'$ portal.}
 In Sec.III we carry out the numerical analysis and find out the parameter space region
 {in which we can satisfy all the relevant experimental constraints. 
 In Sec.IV we discuss two miscellaneous issues for completeness:  (i) breaking of extra $U(1)_{\mu - \tau}$ and 
 (ii) the spin-flipped case where $SU(2)_L$ doublet vectorlike fermions are replaced by colored scalar fields 
 and DM is  $SU(2)$ singlet colorless Dirac fermion. 
 Finally Sec. V is devoted to the summary of our results and the conclusion.}

\section{Model setup and Constraints}
\begin{table}
\begin{tabular}{|c||c|c|}\hline\hline  
& ~$Q'_a$~  & ~$\chi$~ \\\hline\hline 
$SU(3)_C$ & $\bm{3}$  & $\bm{1}$  \\\hline 
$SU(2)_L$ & $\bm{2}$  & $\bm{1}$  \\\hline 
$U(1)_Y$   & $\frac16$ & $0$  \\\hline
$U(1)_{\mu-\tau}$   & $q_x$ & $q_x$  \\\hline
\end{tabular}
\caption{ 
Charge assignments of the new fields $Q'$ and $\chi$  
under $SU(3)_C\times SU(2)_L\times U(1)_Y\times U(1)_{\mu-\tau}$ with $q_x\neq0$ { where we assume these fields have $Z_2$ odd parity.}
Here $Q'$ is vector-like fermions, and its lower index $a$ is the number of family that runs over $1-3$. 
$\chi$ is a complex boson that is considered as a DM candidate.}
\label{tab:1}
\end{table}

\begin{figure}[t]
\begin{center}
\includegraphics[width=70mm]{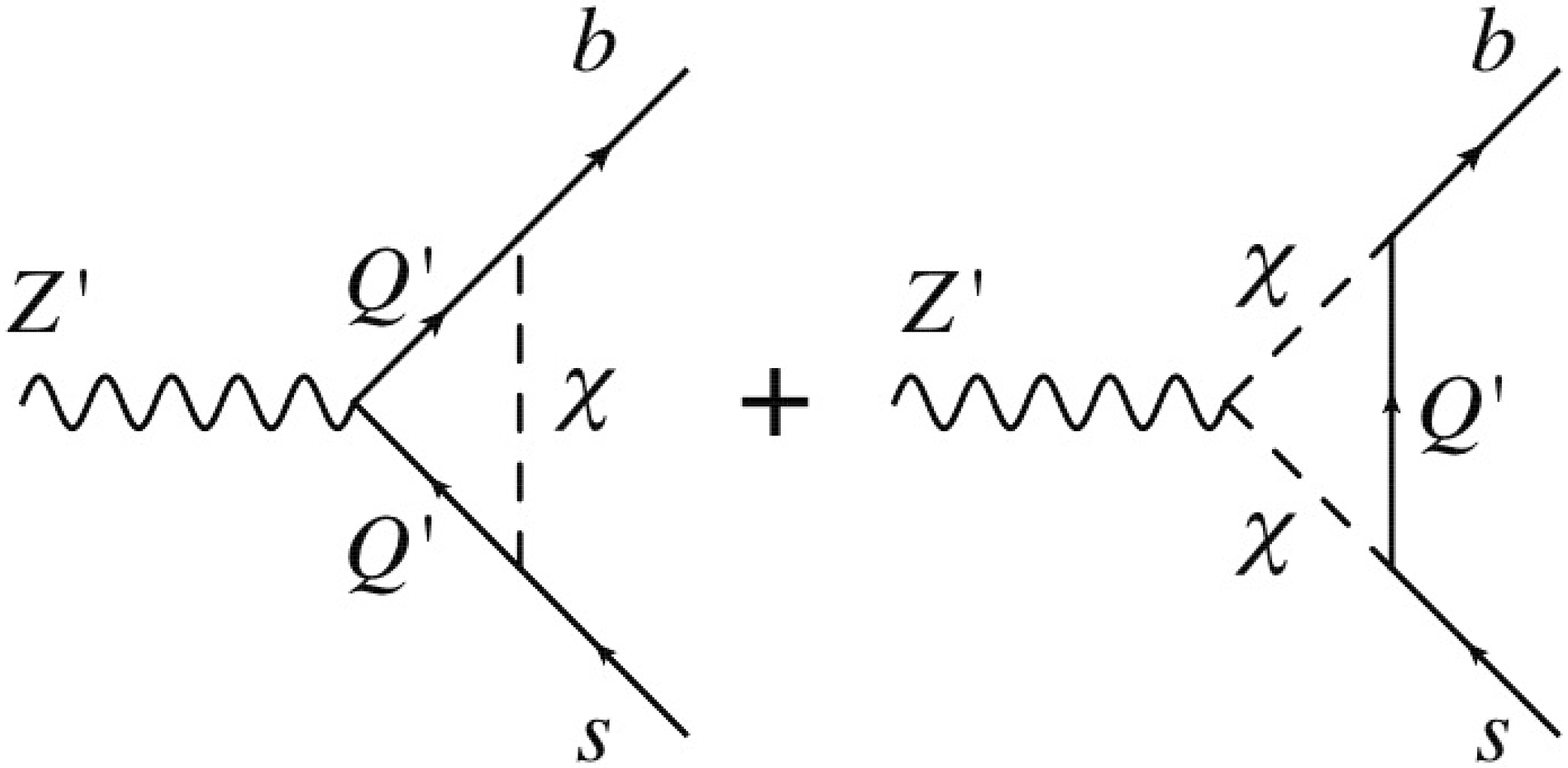} \qquad
\caption{The diagrams introducing effective coupling for $Z'_\mu \bar b \gamma^\mu s + h.c.$ interaction. } 
  \label{fig:diagram}
\end{center}\end{figure}

In this section we set up our model and derive some formula {in $B$ physics and DM 
phenomenology, which will be used in Sec. III for the numerical analysis.}
We introduce three vector-like exotic quarks $Q'$ and a complex scalar boson $\chi$, both of which carry  
nonzero $\mu-\tau$ charges { and odd parity under discrete $Z_2$ symmetry that stabilizes DM}.
Here $\chi$ is the lightest $Z_2$-odd particle, and considered as a DM candidate. 
Charge assignments of these new field are summarized in Table~\ref{tab:1}.

The relevant Lagrangian under these symmetries is given by 
\begin{align}
-\mathcal{L}_{\rm VLQ + \chi}
=&  M_a \bar Q'_a Q' + m_\chi^2 \chi^\dagger \chi 
 +( f_{aj} \overline{Q'_{R_a}} Q_{L_j}\chi + {\rm h.c.}),
\label{Eq:lag-flavor}
\end{align}
where $(a,j)=1-3$ are generation indices, {$Q_{Lj}$'s are the SM quark doublets.  
We have omitted kinetic term and scalar potential associated with $\chi$ for simplicity.}  

The anomaly in $B \to K^{(*)} \ell^+ \ell^-$ decay can be explained by the shift of the Wilson coefficient $C_9$ associated with the corresponding operator $(\bar s\gamma_\mu P_L b)(\mu\gamma^\mu \mu)$.
The effective coupling for $Z'_\mu \bar b \gamma^\mu P_L s + h.c.$ is induced at one loop level as shown in Fig.~\ref{fig:diagram} with the Yukawa coupling in Eq.~(\ref{Eq:lag-flavor}).
Then the effective Hamiltonian $(\bar s\gamma_\mu P_L b)(\bar{\mu}\gamma^\mu \mu)$ arises from $Z'$ mediation 
and the contribution to Wilson coefficient $\Delta C_9^{\mu \mu}$ is obtained as:
\begin{align}
& \Delta C_9^{\mu \mu} \simeq \frac{q_x g'^2}{m_{Z'}^2 C_{\rm SM}}\sum_{a=1-3}f^\dag_{3a}f_{a2}\int[dX]\ln\left(\frac{\Delta[M_a,m_\chi]}{\Delta[m_\chi,M_a]}\right), \nn \\
\label{eq:c9}
& C_{\rm SM}\equiv \frac{V_{tb}V^*_{ts} G_{\rm F}\alpha_{\rm em}}{\sqrt2 \pi},\\
& \Delta[m_1,m_2]=(X+Y-1)(X m_b^2+Y m_s^2) \nn \\
& \qquad \qquad \qquad +X m_1^2 +(Y+Z) m_{2}^2,\nn
\end{align}
where 
$V_{tb}\approx0.999$, $V_{ts}\approx-0.040$ are  the 3-3 and 3-2 elements of CKM matrix respectively,
$G_{\rm F}\approx 1.17\times 10^{-5}$ GeV is the Fermi constant, $\alpha_{\rm em}\approx1/137$ is the 
electromagnetic fine-structure constant, 
$\int_0^1 [dX]\equiv \int_0^1dXdYdZ\delta(1-X-Y-Z)$, 
{$m_b\approx 4.18$ GeV and $m_s\approx {0.095}$ GeV are respectively the bottom and strange quark masses 
given in the $\overline{MS}$ scheme at a renormalization scale $\mu = 2$ GeV~\cite{pdg},}
$m_\chi$ is the mass of $\chi$, and $M_a$ is the mass of $Q'_a$. Notice here that we have assumed 
$m_b,m_s \ll m_{Z'}$ to derive the formula of $C_9$ in Eq.~(\ref{eq:c9}).
The global fit for the value of $ C_9$~\cite{Hurth:2016fbr,Descotes-Genon:2015uva} based on LHCb data suggests that the best fit value is
\begin{equation}
\Delta C_9 \sim - 1. \label{eq:C9_fit}
\end{equation}
In the following numerical analysis, we explore possible value of the $\Delta C_9$ in the model defined in Table~I.

{\it $M-\overline M$ mixing}:   The exotic vector-like quarks and the complex scalar DM $\chi$ induce 
the neutral meson ($M$)-antimeson ($\overline M$) mixings such as $K^0-\bar K^0$, $B_d-\bar B_d$, 
$B_s-\bar B_s$, and  $D^0-\bar D^0$ from the box type one-loop diagrams.  
The formulae for the mass splitting are  respectively given by~\cite{Gabbiani:1996hi}
\begin{align}
& \Delta m_K \approx
\sum_{a,b=1}^3
f^\dag_{1a} f_{a1} f^\dag_{2b} f_{b2} G^K_{box}[m_\chi,M_a,M_b] \nn \\
& \quad \quad \ \,  \lesssim 3.48\times10^{-15} \ [{\rm GeV}],
\label{eq:kk}\\
& \Delta m_{B_d} \approx
\sum_{a,b=1}^3
f^\dag_{1a} f_{a1} f^\dag_{3b} f_{b3} G^{B_d}_{box}[m_\chi,M_a,M_b] \nn \\
& \quad \quad \ \, \lesssim 3.36\times10^{-13} \ [{\rm GeV}],\\
& \Delta m_{B_s} \approx
\sum_{a,b=1}^3
f^\dag_{2a} f_{a2} f^\dag_{3b} f_{b3} G^{B_s}_{box}[m_\chi,M_a,M_b] \nn \\
& \quad \quad \ \, \lesssim 1.17\times10^{-11} \ [{\rm GeV}],\\
& \Delta m_D \approx
\sum_{a,b=1}^3
f^\dag_{2a} f_{a2} f^\dag_{1b} f_{b1} G^{D}_{box}[m_\chi,M_a,M_b] \nn \\
& \quad \quad \ \, \lesssim 6.25\times10^{-15} \ [{\rm GeV}],\label{eq:dd} \\
& G^M_{box}(m_1,m_2,m_3) \nn \\
& \qquad =
\frac{m_M f_M^2}{3(4\pi)^2}
\int_0^1 \frac{X [dX]}{X m^2_1+Y m_2^2+Z m^2_3},\label{eq:mmbar-mix}
\end{align}
where relevant quarks $(q,q')$ are respectively $(d,s)$ for $K$,  $(b,d)$ for ${B_d}$,  $(b,s)$ for ${B_s}$, and  
$(u,c)$ for $D$, each of the last inequalities of the above equations 
represent the upper bound from the experimental values \cite{pdg}, and
$f_K\approx0.156$ GeV, $f_{B_d(B_s)}\approx0.191(0.200)$ GeV, $f_{D}\approx0.212$ GeV,
 $m_K\approx0.498$ GeV,  $m_{B_d(B_s)}\approx5.280(5.367)$ GeV, and  $m_{D}\approx 1.865$ GeV.

 {\it $b\to s\gamma$}:   
 $\Gamma_{b\to s\gamma}$ in our model is given by 
\begin{align}
& \Gamma_{b\to s\gamma} \approx
\frac{\alpha_{\rm em}m_b^3 }{12(4\pi)^4}(m_b^2+m_s^2)
\left|\sum_{a=1}^3
\frac{f^\dag_{2a}f_{3a} F(M_a,m_\chi)}{36(M_a^2-m_\chi^2)^4} \right|^2, \nn \\
&F(m_1,m_2) 
= 5 m_1^6-27 m_1^4 m_2^2+27 m_1^2m_2^4 -5m_2^6 \nn \\
& \qquad \qquad \qquad -12m^4_2(-3m_1^2+m_2^2)\ln(m_1/m_2),
\label{eq:DRbtosg}
\end{align}
then the branching ratio ${\rm BR}(b\to s\gamma)$ and its constraint is found as
\begin{align}
&{\rm BR}(b\to s\gamma)
\equiv \frac{\Gamma(b\to s\gamma)}{\Gamma_{\rm tot.}}\lesssim3.29\times 10^{-4},\\
&{\Gamma}_{\rm tot.}\approx4.02\times10^{-13}\ {\rm GeV}. 
\label{eq:BRbtosg}
\end{align}
 
{\it Constraints from direct production of $Q'$s}: The exotic quarks $Q'$s can be pair produced via QCD
processes at the LHC and then each $Q'$ will decay through $Q' \rightarrow q_i \chi$ where $q_i$ represents 
a quark with flavor $i$.
Therefore search for ``$\{ t t, b b , t  j, b j , j j  \}$ + missing $E_T$" signals will constrain our model, 
the branching ratios into a particular quark flavor $i$ depending on the relative sizes of 
Yukawa couplings, $f_{3j}$ and $f_{aj}$ with $a=1,2$.  
We roughly estimate the lower limit on the mass of $Q'$ from the current LHC data for squark searches~\cite{CMS:2016mwj, Aaboud:2016zdn}, which indicates the mass should be larger than $\sim 0.5$-$1$ TeV depending on the mass difference between $Q'$ and $\chi$. In our following analysis, we simply take $M_{a} > 1$ TeV to satisfy this  constraint.

{\it Dark matter} : 
In our scenario, complex scalar $\chi$ is considered as a DM candidate which dominantly annihilate into SM leptons 
via $\chi \chi \to Z' \to \mu^+ \mu^- (\tau^+ \tau^-)$,~\footnote{Notice here that the cross section mode via 
Yukawa coupling $g$ gives the d-wave suppression. 
Thus the $s$-wave mode via $Z'$ boson is dominant. In this case, however, one cannot apply $v_{\rm rel}$ 
expansion approximation to compute the relic density, since it has the pole solution near $2m_\chi\approx m_{Z'}$. 
Thus we treat it with more precise way~\cite{Nishiwaki:2015iqa, Edsjo:1997bg}. } so that the DM in our model 
is naturally leptophilic.
The relic density of DM is given by
\begin{align}
&\Omega h^2
\approx 
\frac{1.07\times10^9}{\sqrt{g_*(x_f)}M_{Pl} J(x_f)[{\rm GeV}]},
\label{eq:relic-deff1}
\end{align}
where $g^*(x_f\approx25)\approx100$, $M_{Pl}\approx 1.22\times 10^{19}$,
and $J(x_f) (\equiv \int_{x_f}^\infty dx \frac{\langle \sigma v_{\rm rel}\rangle}{x^2})$ is given by
\begin{align}
&
J(x_f)=\int_{x_f}^\infty dx\left[ \frac{\int_{4m_\chi^2}^\infty ds\sqrt{s-4 m_\chi^2} (\sigma v_{\rm rel}) K_1\left(\frac{\sqrt{s}}{m_\chi} x\right)}{16  m_\chi^5 x [K_2(x)]^2}\right], \nn \\
& (\sigma v_{\rm rel})
= 
\frac{g'^4 x^2  s (s-m_\chi^2)}{3\pi(s-m_{Z'}^2)^2}.
\label{eq:relic-deff2}
\end{align}
Here  $s$ is  a Mandelstam variable, and $K_{1,2}$ are the modified Bessel functions of the second kind 
of order 1 and 2, respectively.
In our numerical analysis below, we use the current experimental 
range for the relic density: $0.11\le \Omega h^2\le 0.13$~\cite{Ade:2013zuv}.
Notice here that we simply assume {the Higgs portal coupling for $\chi-\chi-h_{\rm SM}$} 
to be tiny enough to evade the direct detection via Higgs exchange. However we have the process via $Z'$ portal,
and its spin independent scattering cross section  is given by~\cite{Khalil:2011tb}
\begin{align}
&\sigma\approx 
\frac{C_{\rm eff}}{16\pi}\left(\frac{m_\chi m_N}{m_N+m_\chi}\right)^2 \frac{q_x^2 g'^2}{(4\pi)^2m_{Z'}^2} 
\nn \\& \quad \times 
\left| \sum_{a=1}^3 f^\dag_{1a}f_{a1}\int[dX]\ln\left(\frac{\Delta[M_a,m_\chi]}{\Delta[m_\chi,M_a]}\right)\right|^2\ ,
\label{eqq:dd}
\end{align}
where $C_{\rm eff}\approx6.58\times10^{-24}$, and $m_N\approx0.939$ GeV.
The current experimental upper bound is $\sigma_{\rm exp}\lesssim 2.2\times10^{-46}$ cm$^2$ at $m_\chi\approx50$ GeV according to the LUX data~\cite{Akerib:2016vxi}.
In our numerical analysis, we conservatively restrict the LUX bound for the whole the DM mass range.

{\it $U(1)$ kinetic mixing}: The kinetic mixing between $U(1)_Y$ and $U(1)_{\mu -\tau}$ is induced by fermion loops including $\mu$, $\tau$ and $Q'_a$.  The kinetic mixing term is given by 
\begin{equation}
{\cal L}_{\rm mix} =\epsilon/2 B_{\mu \nu} X^{\mu \nu},
\end{equation} 
where $B_{\mu \nu}$ and $X_{\mu \nu}$ are respectively the field strength of $U(1)_Y$ and $U(1)_{\mu-\tau}$ gauge fields. The mixing parameter $\epsilon$ is roughly obtained as 
\begin{equation}
\epsilon \sim \frac{e g'}{6 \pi^2} \ln (m_\tau/m_\mu) + \frac{q_x e g' }{6 \pi^2} \sum_a \ln(\Lambda/M_a)
\end{equation}
 where $\Lambda$ is some heavy scale~\cite{Holdom:1985ag, Dienes:1996zr, Jaeckel:2012yz, Chiang:2013kqa}; for example it can be heavy vector like quark with opposite $U(1)_{\mu -\tau}$ charge. For $g' = 0.1$ the size of mixing parameter is $|\epsilon| \lesssim 10^{-3} $ if $\Lambda$ is not too large compared to $M_a$. 
In such a case, $Z$-$Z'$ mixing angle is roughly given by $\theta_{ZZ'} \sim \epsilon (m_Z^2/m_{Z'}^2)$. Thus the effect of $Z$-$Z'$ mixing is small in decays of $Z$ boson and SM fermions due to small mixing angle; the mixing is also consistent with other constraints~\cite{Williams:2011qb}.

\section{Numerical analysis \label{sec:numerical}}
\begin{figure}[t]
\begin{center}
\includegraphics[width=60.0mm]{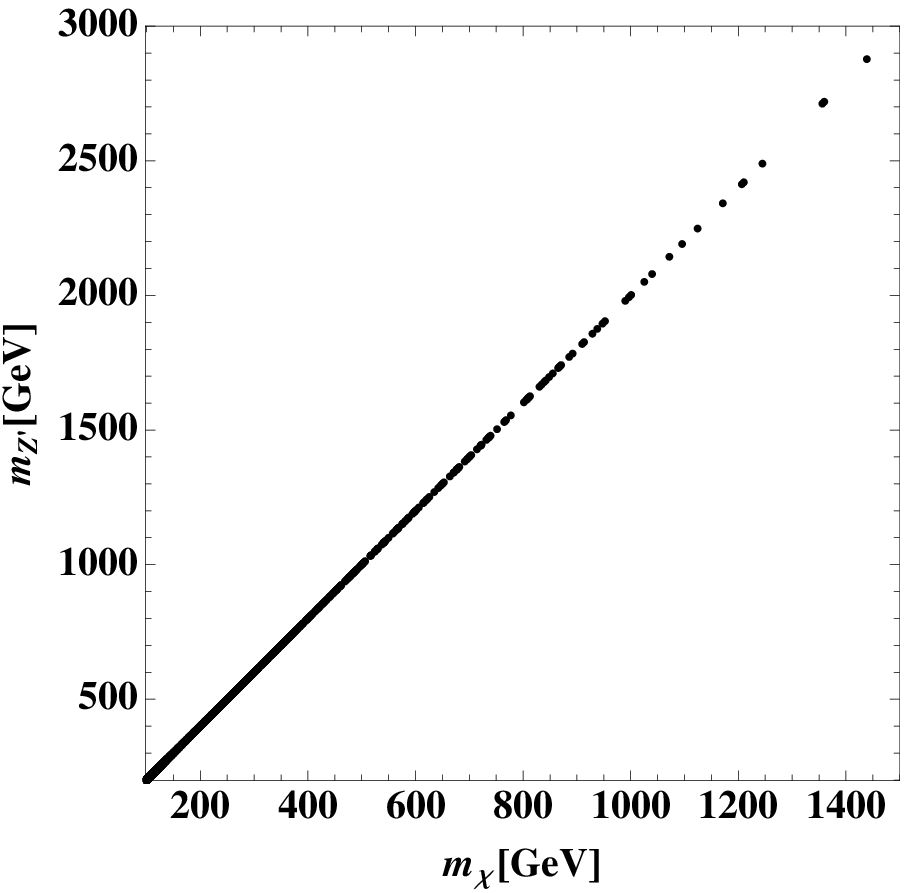} \qquad
\includegraphics[width=65.0mm]{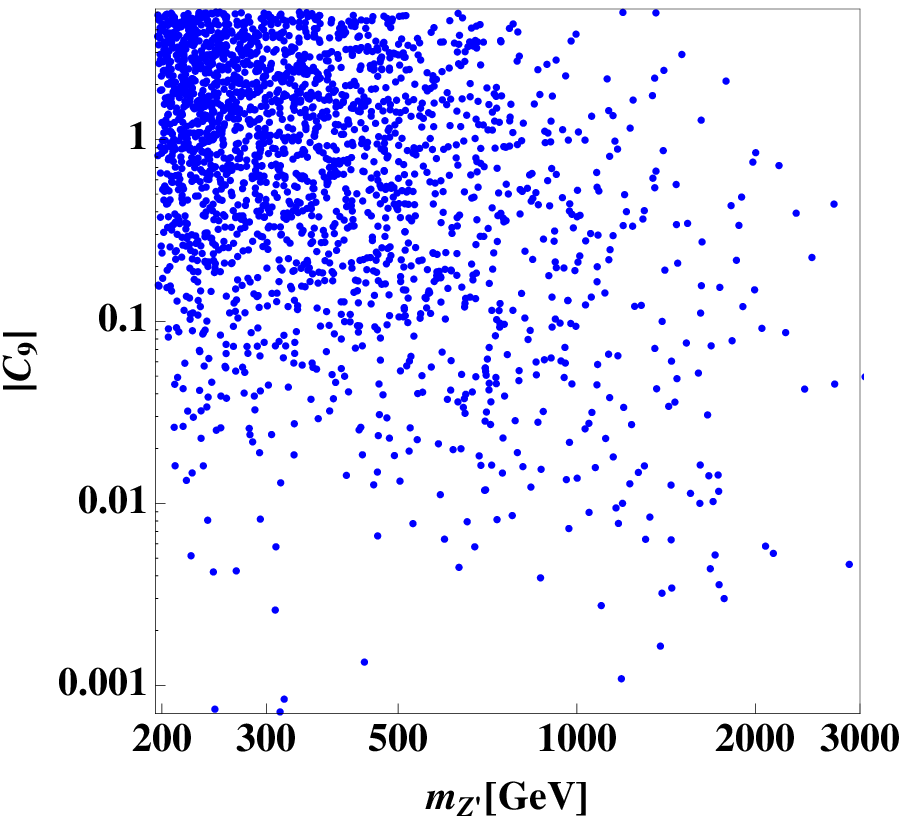}
\caption{ The top panel represents the allowed range for $m_\chi$ and $m_{Z'}$, while the bottom one does 
the allowed range for $m_{Z'}$ and $\Delta C_9$. The correlation between $m_\chi$ and $m_{Z'}$ in the top panel 
arises from the relation of relic density of DM. In the bottom panel, one finds that one can obtain 
$\Delta C_9 \sim -1$ for $m_{Z'} \lesssim 2$ TeV that is required to resolve $B \to K^{(*)} \ell^+ \ell^-$ anomalies. 
} 
  \label{fig:DM-x-Zp}
\end{center}\end{figure}

In this section, we perform the numerical analysis.
First of all, we fix two parameters $g'=0.1$ and $|q_x|=1$ for simplicity.
In this case, the lower bound on the mass of $Z'$ is at about $60$ GeV, which arises from the neutrino trident production~\cite{Altmannshofer:2014pba}. {On the other hand, the effective operator to obtain $\Delta C_9 \sim -1$} requires rather large $Z'$ mass~\footnote{Here we set the lowest bound on $Z'$ mass as $m_{Z'} \geq$ 200 GeV.}. 
Thus this bound is  always safe in our case.
The ranges of the other input parameters are set to be as follows: 
\begin{align}
& 
f  \in[10^{-3},1],\  m_{Z'}\in [{ 200}, 3000]\ [{\rm GeV}],\ \nn \\
& m_{\chi} \in [1,2000]\ [{\rm GeV}], \ M_{a}\in [{ 1000},3000]\ [{\rm GeV}] . 
\end{align}
We also assume $M_1 < M_2 < M_3$, $m_{Z'} > m_\chi$, and take $m_\chi < 1.2 M_1$   for simplicity so that we can 
ignore contributions from coannihilation processes. 
Then we randomly scan over $3 \times 10^7$ parameter points in the above ranges 
and select the points that satisfy all the constraints such as {\it $M-\overline M$ mixing, $b\to s\gamma$ branching 
ratio, measured relic density of DM, the spin independent DM-nucleon scattering cross section via $Z'$ portal} as discussed 
in the previous section. 
In the left panel of Figs.~\ref{fig:DM-x-Zp}, we show the allowed parameter region for $m_\chi$ and $m_{Z'}$.
The correlation between $m_\chi$ and $m_{Z'}$ in this plot arises from the relation of relic density of DM,
which indicate the relation $m_{Z'} \sim 2 m_\chi$ is required to obtain the relevant DM annihilation cross 
section via the $s$-channel resonant enhancement. 
On the other hand, the right panel of Fig.~\ref{fig:DM-x-Zp} represents the allowed range for $m_{Z'}$ 
and $\Delta C_9$.  In this plot, one can easily obtain $\Delta C_9 \sim -1$   for $m_{Z'} \lesssim 2000$ GeV 
so that one can resolve $B \to K^{(*)} \ell^+ \ell^-$ anomalies. 
Notice here that the most stringent bound on $C_9$ arises from the constraint of $B_{s}-\bar B_{s}$ mixing.

\section{Miscellaneous Issues}
\subsection{Effects of $U(1)_{\mu-\tau}$ symmetry breaking on $B$ physics}

It is worthwhile to mention that the $U(1)_{\mu-\tau}$ breaking mechanism
does not affect the $B$ physics anomalies that is our main subject. 
Here let us for example consider the singlet scalar $\phi$ with charge $2$, and 
assume $\chi$ has charge 1 for simplicity. Then there is a term (dim-3) $\chi^2 \phi^\dagger + H.c.$ which 
breaks $U(1)_{\mu-\tau}$ into $Z_2$ subgroup, $\chi \rightarrow - \chi$ in the scalar potential.
In this framework neutrino masses and their mixings can be fitted to the current experimental data~\cite{Heeck:2011wj}.
As for such kind of model, see Ref.~\cite{Baek:2014kna} in the dark U(1) case.
Also one finds the $Z_3$ case in Ref.~\cite{Ko:2014nha} , if we choose the $\phi$ charge is 3.
In this case an additional contribution to the relic density of DM and the direct detection via Higgs portal are arisen,
and we can relax the resonant allowed region in the left panel of Fig.~\ref{fig:DM-x-Zp}.
 
 \subsection{Variation where the new fields spins are flipped}
 Here let us briefly mention on a variation of our model where new particle spins are 
 flipped: namely we consider  $SU(2)_L$ doublet colored scalars and a gauge singlet Dirac fermion,  
 like SUSY partners. Let us define $\tilde Q'$ as the $SU(2)_L$ doublet scalar boson, and  $\tilde\chi$ as the gauge singlet (Dirac) fermion. Then one finds a Yukawa Lagrangian $f'_{ij}\bar{Q}_{L_i} \tilde\chi_{R_j} \tilde {Q'}+{\rm h.c.} $.~\footnote{Notice here the sign of $U(1)_{\mu-\tau}$ charge assignments between $\tilde \chi$ and $\tilde Q'$ are taken to be opposite, although the the absolute value is the same.} Even in this case, the result for the $\Delta C_9$ 
 is almost the same as one in the original model setup. However a remarkable difference arises in the relic density 
 of DM, where the Dirac fermion $\tilde\chi$ is considered as the DM candidate. Then its annihilation cross section, 
which is $s$-wave dominant, is given by 
\[
\sigma v_{\rm rel}\approx \frac{m_{\tilde\chi}^2}{32\pi(m_{\tilde\chi}^2+m_{\tilde Q'}^2)^2} + \frac{g'^4 x^2 m_{\tilde\chi}^2}{16\pi(-4 m_{\tilde\chi}^2+m_{Z'}^2)^2}
\] 
even in the limit of the massless final state. Then it suggests that the allowed region that satisfies thermal relic 
density would be  wider.  The $M-\overline M$ mixing, which arises from $Q_1$ operator~\cite{Gabbiani:1996hi}, 
is the same as one in the original model. Therefore it does not give stringent constraints.

Furthermore, an interesting phenomenology will appear if $|q_x|=1$.
In this case, one has an additional term 
\[
g'_{ij}\bar d_{R_i} L_{L_j} i\sigma_2 \tilde Q'^T +{\rm h.c.} , 
\]
that would induces the following operators:
\[
\frac{g'_{b\ell}g'_{s\ell'}}{4 m_{Q'^2}}(\bar s\gamma^\mu P_R b)(\ell'\gamma_\mu\ell)~~~
{\rm and}~~~  -\frac{g'_{b\ell}g'_{s\ell'}}{4 m_{Q'^2}}(\bar s\gamma^\mu P_R b)(\ell'\gamma_\mu\ell) , 
\] 
which respectively correspond to $C_9'$ and $C'_{10}$ with $C'_9=-C'_{10}$.
These Wilson coefficients can also  resolve anomalies in $B \to K^{(*)} \ell^+ \ell^-$  
decay~\cite{Descotes-Genon:2015uva}. 
Notice here that for $|q_x|=1$ the colored scalar {$\tilde{Q}'$} is identical to a scalar leptoquark. 
Therefore its mass is strongly constrained by the LHC data that its lower bound is about 1 TeV as in the 
vector like quark case discussed in the previous section.

\subsection{$Z'$ production at the LHC}
The $Z'$ boson can be produced at the LHC via loop induced couplings to SM quarks, $g_{\rm eff} (\bar q \gamma^\mu P_L q') Z'_\mu$.
Here we consider the case where $\Delta C_9 \sim -1$ is obtained by fixing extra $U(1)$ gauge coupling and charge for $\{ Q'_a, \chi \}$ as $g' = 0.1$ and $|q_x|=1$.
In this case, the $g_{\rm eff} \sim 0.002$ is required for $q(q') = s (b)$ for $m_{Z'} = 500$ GeV. 
Then we consider two scenarios for illustration: (i) $g_{\rm eff}  = 0.002$ for all quark combinations; (ii) $g_{\rm eff} = 0.002$ for operators including only second and third generation quarks but $g_{\rm eff} = 0$ if first generation quarks are included, where $m_{Z'} =500$ GeV is fixed for both scenarios.
The $Z'$ production cross section is estimated with {\it CalcHEP}~\cite{Belyaev:2012qa} by implementing relevant interactions and using $\sqrt{s} = 13$ TeV.
We obtain the cross section such as $\sigma_{pp \to Z'} \simeq 1.6 \times 10^{-2}[1.1 \times 10^{-3}]$ pb for scenario (i)[(ii)].
Assuming $m_{Z'} < 2 m_\chi (2 M_a)$, the dominant branching ratio for $Z'$ decay is given by $BR(Z' \to \mu^+ \mu^-) \simeq BR(Z' \to \tau^+ \tau^-) \simeq 0.5$. 
Then we find the cross section for scenario (i) is marginal of the current upper limit by the LHC data from $Z' \to \mu^+ \mu^-$ search~\cite{ATLAS:2016cyf, CMS:2016abv} while that of scenario (ii) is much lower than the current limit. These cross sections will be further tested by data with more integrated luminosity. 

\section{Summary and Conclusions}
In this paper, we have proposed an extension of the SM with three families of  exotic quarks and
an inert singlet scalar boson $\chi$ imposing a gauged $\mu-\tau$ symmetry.
Then we have explained the measured anomalies in $B \to K^{(*)} \ell^+ \ell^-$ through the one-loop radiative effect
and relic density of dark matter $\chi$ without conflict with the constraints from spin independent dark matter direct detection searches via $Z'$ boson exchange, $M-\overline M$ mixing processes, and branching ratio of $b\to s\gamma$.

We have shown the allowed parameter region that is consistent with all the relevant constraints. 
The left panel of Figs.~\ref{fig:DM-x-Zp} shows the allowed range for $m_\chi$ and $m_{Z'}$, in which
we have shown the correlation between $m_\chi$ and $m_{Z'}$ that arises from the relation of relic density of DM,
indicating $m_{Z'} \sim 2 m_\chi$ to enhance the annihilation cross section through the $s$-channel resonance.
On the other hand we  have shown  the allowed range for $m_{Z'}$ and $\Delta C_9$ in  the right panel of 
Figs.~\ref{fig:DM-x-Zp}.
In this figure, we have found that one can easily obtain $\Delta C_9 \sim -1$ for $m_{Z'} \lesssim 2000$ GeV, 
resolving the $B \to K^{(*)} \ell^+ \ell^-$ anomalies. 

In addition, we have discussed the case 
where the extra particle spins are flipped.   
In this variational model, we could obtain the required $\Delta C_9$ in a similar manner, while dark matter 
annihilation cross section is $s$-wave dominant so that  the allowed parameter region is extended.
Note that other constraints are similar to the original model setup.   The case $|q_x|=1$ is special since in this case 
$\tilde{Q}'$ becomes a scalar leptoquark. Then we have additional contributions to $b\rightarrow s \ell^+ \ell^-$ in such 
a way $C_9' = -C_{10}'$  that  also help to resolve the $B \to K^{(*)} \ell^+ \ell^-$ anomalies.

Before closing, we emphasize that our mechanism of generating flavor violating $Z'$ couplings 
can be generalized readily by including both quark and lepton sectors by selecting the $Z_2$ odd exotic particle 
contents. Therefore this mechanism provides interesting connection between flavor physics and dark matter physics 
where our model represents one explicit example giving connection between $B$-physics and dark matter physics.


\section*{Acknowledgments}
H. O. is sincerely grateful for all the KIAS members, Korean cordial persons, foods, culture, weather, and all the other things.
This work is supported in part by National Research Foundation of Korea (NRF) Research Grant 
NRF-2015R1A2A1A05001869 (PK), and by the NRF grant funded by the Korea government (MSIP) 
(No. 2009-0083526) through Korea Neutrino Research Center at Seoul National University (PK). 

\end{document}